\documentclass[10pt,letterpaper]{article}
\usepackage[top=0.85in,left=2.75in,footskip=0.75in]{geometry}
\usepackage{caption}
\usepackage{subcaption}
\usepackage{amsmath,amssymb}

\usepackage{changepage}

\usepackage{textcomp,marvosym}

\usepackage{cite}

\usepackage{nameref,hyperref}

\usepackage[right]{lineno}

\usepackage[nopatch=eqnum]{microtype}
\DisableLigatures[f]{encoding = *, family = * }

\usepackage[table]{xcolor}
\usepackage{makeidx}  
\usepackage{ifpdf}
\usepackage{url}
\usepackage{graphicx} 
\usepackage{array}

\usepackage[normalem]{ulem}
\useunder{\uline}{\ul}{}

\usepackage{tablefootnote}

\newcolumntype{+}{!{\vrule width 2pt}}

\newlength\savedwidth

\newcommand\thickhline{\noalign{\global\savedwidth\arrayrulewidth\global\arrayrulewidth 2pt}%
\hline
\noalign{\global\arrayrulewidth\savedwidth}}

\raggedright
\usepackage[aboveskip=1pt,labelfont=bf,labelsep=period,justification=raggedright,singlelinecheck=off]{caption}

\makeatletter
\renewcommand{\@biblabel}[1]{\quad#1.}
\makeatother

\usepackage{setspace}

\usepackage{notes2bib}
\usepackage{lastpage,fancyhdr,graphicx}
\usepackage{epstopdf}
\fancyheadoffset[L]{1in}
\fancyfootoffset[L]{0in}
\begin{document}
\vspace*{0.2in}
\doublespacing
\begin{flushleft}
{\Large
\textbf\newline{Machine Learning Driven Biomarker Selection for Medical Diagnosis} 
}
\newline
\\
Divyagna Bavikadi\textsuperscript{1*},
Ayushi Agarwal\textsuperscript{1},
Shashank Ganta\textsuperscript{1},
Yunro Chung\textsuperscript{2,3},
Lusheng Song\textsuperscript{2},
Ji Qiu\textsuperscript{2},
Paulo Shakarian\textsuperscript{1}
\\
\bigskip
\textbf{1}  Fulton Schools of Engineering, Arizona State University, Tempe, AZ, USA
\\ 
\textbf{2} Biodesign Center for Personalized Diagnostics,  Arizona State University, Tempe, AZ, USA
\\
\textbf{3} College of Health Solutions, Arizona State University, Phoenix, AZ, USA
\\
\bigskip

%
%



* dbavikad@asu.edu 

\end{flushleft}
\section*{Abstract}
Recent advances in experimental methods have enabled researchers to collect data on thousands of analytes simultaneously.  This has led to correlational studies that associated molecular measurements with diseases such as Alzheimer's, Liver, and Gastric Cancer. However, the use of thousands of biomarkers selected from the analytes is not practical for real-world medical diagnosis and is likely undesirable due to potentially formed spurious correlations.  In this study, we evaluate $4$ different methods for biomarker selection and $4$ different machine learning (ML) classifiers for identifying correlations – evaluating $16$ approaches in all. We found that contemporary methods outperform previously reported logistic regression in cases where $3$ and $10$ biomarkers are permitted.  When specificity is fixed at $0.9$, ML approaches produced a sensitivity of $0.240$ ($3$ biomarkers) and $0.520$ ($10$ biomarkers), while standard logistic regression provided a sensitivity of $0.000$ ($3$ biomarkers) and $0.040$ ($10$ biomarkers).  We also noted that causal-based methods for biomarker selection proved to be the most performant when fewer biomarkers were permitted, while univariate feature selection was the most performant when a greater number of biomarkers were permitted.



\section*{Introduction}
Recent advances in experimental methods have enabled researchers to collect data on thousands of analytes (biological analytes) simultaneously (Rosado et al.~\cite{bib1}, Topkaya et al.~\cite{bib2}).  This has led to correlational studies that associated these molecular measurements with diseases such as Alzheimer's (Blennow et al.~\cite{bib3}), Liver (Ahn Joseph C et al.~\cite{bib4}), and Gastric Cancer (Lin et al.~\cite{bib5}).  However, it is generally considered undesirable to use thousands of biomarkers selected from the analytes for medical diagnosis for several reasons.  First, large numbers of biomarkers increase the likelihood of spurious correlation.  Second, the use of many biomarkers increases model complexity and hinders the interpretability of results.  Further, from a practical standpoint, the use of fewer biomarkers is preferable from the standpoint of creating cost-effective diagnostic products.
\\
As a result, previous studies have conducted two operations in tandem: the selection of candidate biomarkers thought to be associated with a given disease individually and the identification of correlations between the combination of selected candidate biomarkers and the target medical condition.  The most commonly reported methodology in the literature has been logistic regression, often accompanied by a variant of univariate feature selection (Bursac et al.~\cite{bib6}, Direkvand-Moghadam et al.~\cite{bib7}, Islam et al.~\cite{bib8}).  This paper looks to augment existing work by studying the effect of the feature selection method and model type.  In particular, we examine causal-based feature selection (Kleinberg et al.~\cite{bib9}) and a variety of machine-learning approaches, including gradient-boosted decision trees and neural networks.  In all, we study $16$ different combinations of feature selection and classification models in tests where the number of biomarkers $K$ is restricted to a set of values $1,3,4,10,15,30$ on a gastric cancer dataset that includes measurements from $3440$ biological analytes (Song et al.~\cite{bib10}).  We perform a cross-validation study and report results on training and test sets as well as examine hyperparameter sensitivity for the causal-based approaches.  We found that contemporary machine learning methods outperform previously reported logistic regression in these experiments.  When specificity is fixed at $0.9$, ML approaches produced a sensitivity of $0.240$ (3 biomarkers) and $0.520$ (10 biomarkers), while standard logistic regression provided a sensitivity of $0.000$ (3 biomarkers) and $0.040$ (10 biomarkers).
\\
The rest of the paper is organized as follows:  We first provide a brief overview of related work, a description of the gastric cancer dataset, and machine learning methods.  This is followed by reporting of the experimental results on the gastric cancer dataset and associated discussion. Finally, we conclude by discussing our findings.

\section*{Related Work} 
Machine learning models, such as logistic regression, have been utilized with biological data for association purposes. In (Islam et al.~\cite{bib8}), the correlation coefficients of three biomarkers: body temperature, heart rate, and probable blood glucose level, were evaluated and associated with malaria detection using logistic regression. Similarly, in (Direkvand-Moghadam et al.~\cite{bib7}), univariate logistic regression demonstrated a substantial association between female sexual dysfunction and biomarkers, such as age, gravidity, and menarche age. Additionally, in (Bursac et al.~\cite{bib6}), the application of feature selection prior to model training showed the potential to maintain confounding variables, especially when dealing with macro biological data sets. Note that none of this prior work conducts an analysis of various machine learning classifiers, such as gradient-boosted trees or neural networks with causal-based and feature selection methods.
\\
More specifically, machine learning models paired with feature selection for disease detection have proved significantly beneficial. In (Sorino et al.~\cite{bib11}), numerous machine learning techniques similar to ours, such as random forest classifier and boosted tree classifier, with cross-validation were used to diagnose non-alcoholic fatty liver disease. Similarly, in (Díaz Álvarez et al.~\cite{bib12}), a feature selection, evaluated on chi-squared statistic was paired with a Naive Bayes classifier to aid the diagnosis and classification of neuro-degenerative disorders. Moreover, vision-based machine learning techniques such as convolutional neural networks have been applied to a wide variety of medical diagnostic use cases (Yadav et al.~\cite{bib13}, Shaban et al.~\cite{bib14}, Heenaye-Mamode et al.~\cite{bib15}, Lopez-Garnier et al.~\cite{bib16}, Kundu et al.~\cite{bib17}). Such diagnosis based on imagery would be complementary to biomarker-based diagnosis. However, to our knowledge, the application of such techniques to the use of biomarkers, specifically proteins, for the purposes of medical diagnosis has not been studied in the literature.
The concept of causal-based methods, such as the one apparent in our findings, has been used in a variety of medical applications (Kleinberg et al.~\cite{bib18}). For example, in (Richens et al.~\cite{bib19}), the application of causal machine learning effectively increased clinical accuracy from the top $48\%$ to the top $25\%$ of doctors. However, to date, such methods have not been combined with recent advances in biomarker experimentation (Kleinbaum et al.~\cite{bib20}) for medical diagnosis based on biomarker measurements.

\section*{Gastric Cancer Dataset}
The dataset (Song et al.~\cite{bib21}) used for the biomarker discovery contains information on $100$ samples, each of which is associated with a case or control indicating the presence or absence of gastric cancer. The dataset is balanced with $50$ samples labeled case and $50$ samples labeled control. The age and gender of the samples are matched between cases and controls. 
Each instance is represented by $3440$ corresponding molecular measurement values, which are used to assess 
the risk of gastric cancer and provide insight into the disease. The measurement values range from $0.00$ to $260.65$. Molecular measurements were noted with IgG and IgA antibodies against the same set of proteins. The dataset contains data on clinical features, antibody reactions against \textit{Helicobacter pylori} proteins, and demographic variables. Using the Nucleic Acid Programmable Protein Array (NAPPA) technology, the study assessed humoral responses to $1527$ proteins or almost the whole \textit{H. pylori} proteome. The total set of proteins nearly composes a complete \textit{H. pylori} proteome. Measurement values were assessed on seropositivity. Seropositivity was defined as the median normalized intensity $ 2 \leq $ on NAPPA. Table~\ref{table1} shows the breakdown of the dataset.

\begin{table}[!ht]
\centering
\caption{\textbf{Breakdown of Gastric Cancer Dataset}}
\begin{tabular}{|l|l+l|l|l|} 
\hline
\textbf{Data} & \textbf{Samples} & \textbf{Analytes Data} & \textbf{Quantity}   \\ 
\thickhline
Total Samples & $100$ & Total measurements* & $3440$ \\
Cancer Cases & $50$ & Organism: H. Pylori & $3054$ \\
Cancer Controls & $50$ & Organism: EBV & $178$ \\
   &   & Organism: Streptococcus\_gallolyticus & $92$ \\ 
   &   & Organism: Fusobacterium\_nucleatum & $84$ \\ 
   &   & Organism: Other ($\le$5 occurrences) & $32$ \\
   [1ex] 
\hline
\end{tabular}
\label{table1}
\caption*{* indicates that it includes $IgG$ and $IgA$ antibodies}
\end{table}
For the training data, each sample has a vector of real values associated with each analyte measurement and a ground truth that indicates the actual presence of the disease to distinguish between gastric cancer patients from healthy controls.

\section*{Machine Learning and Feature Selection Methods}
\subsection*{Overview of approaches}  
We employ a two-step process for each method: feature selection and classification, and will discuss each in turn.  We will use the symbol $K$ (let $K\in \{1,3,4, 10,15,30\}$) to denote the maximum number of biomarkers permitted after the feature selection step. The best $K$ biomarkers are used to then classify a sample.  We also explore the effect of binarizing biomarker inputs – the intuition being that rather than considering the biomarker measurement directly, we only consider if the biomarker exceeds some threshold $\gamma$ ($\gamma\in \{0.6, 1.0, 1.4, 1.8\}$), which is specified as a hyperparameter.
\\
\subsection*{Feature selection methods}
We consider two types of feature selection methods: the univariate selection and the causal metric.  Univariate feature selection evaluates the strength of the relationship between the feature and the response variable. In this paper, we use chi-square statistic-based univariate feature selection method. By contrast, the causal-based method examines the effect of a single analyte based on other analytes that may have a co-occurring measurement. 
A contribution of this work is an adaption of the causal measure of (Kleinberg et al.~\cite{bib18}) for biomarker selection.  While (Kleinberg et al.~\cite{bib18})  computes causality as the average increase in the probability of the effect when the cause is present, here we propose a new metric based on the intuition of  (Gardner et al.~\cite{bib22}) but adapted for biomarker selection as follows:
\begin{eqnarray}
\label{eqn:metric}
causal(i)=\frac{\sum_{j \in R_i}f(i,j)-f(\neg i, j)}{size(R_i)} 
\end{eqnarray}

Here we still examine the average increase of a function when the biomarker is present based on co-occurring biomarkers.  However, unlike in (Kleinberg et al.~\cite{bib18}) we do not use probability, but a measure more tuned to our domain. In Equation~\ref{eqn:metric} the symbol $causal(i)$ is the causal metric for the analyte $i$, $R_i$ indicates the set of analytes that are related to the analyte $i$, and $f$ indicates the measure calculated based on the product of sensitivity and specificity for every pair of a analyte $i$ and its related analyte $j$.
This makes it more suitable for the kind of protein biomarkers used from the dataset.
We provide details as to how we derived this measure in the supporting information. 

\subsection*{Machine learning classification methods}
We examine four machine learning methods: logistic regression, random forest, deep neural networks (DNN), gradient-boosted decision trees (Pedregosa et al.~\cite{bib23}), and XGBoost (Chen et al.~\cite{bib24}).  The intuition for using logistic regression is to establish it as a baseline as it was used in previous biomarker studies (Direkvand-Moghadam et al.~\cite{bib7}, Ravi et al.~\cite{bib25}), random forest for its ability to provide accurate results with minimal hyper-parameter tuning, a DNN due to their state-of-the-art performance in a variety of other tasks, and two variants of boosted trees which have been shown to provide state-of-the-art performance on tasks involving tabular data.  For the DNN, we employ a dense, multi-layer perceptron with 4 layers, RELU activation function, and a softmax output layer using the PyTorch (Paszke et al.~\cite{bib26}) software package.  For the boosted decision trees, we use the Scikit-learn implementation of gradient-boosted trees and the standard implementation of XGBoost. Summaries of these methods, along with hyperparameter settings can be found in the supporting information.

\section*{Results}
\subsection*{Setup}
We conducted experiments using an NVIDIA GTX1080 (2560 cuda cores, 10 Gbps memory speed).  For evaluation, we used leave-one-out cross-validation (LOOCV) and examined values for Area Under the Curve (AUC) for both training and test data, as well as sensitivity on the test data with specificity fixed at $0.8$ and $0.9$ (sensitivity at specificity of $0.8$ ($Sen@80$) and sensitivity at specificity of $0.9$ ($Sen@90$)). These metrics are selected based on standards employed in assessing diagnostic biomarkers; it also helps us have an overall understanding of performance across multiple confidence thresholds as well as judge the degree to which the model can discriminate between case and control. Evaluation of experiments is conducted on this standard based on other factors such as models, and hyperparameters.
Throughout the discussion, we will treat logistic regression with univariate selection as the baseline, as logistic regression was employed in prior work (Direkvand-Moghadam et al.~\cite{bib7}, Islam et al.~\cite{bib8}).

\subsection*{Selection of 3 Biomarkers}
\begin{figure}[!ht]
\caption{{\bf ROC Curve for XGB model with causality measure (3 Biomarkers).}
}
\includegraphics[scale=0.6]{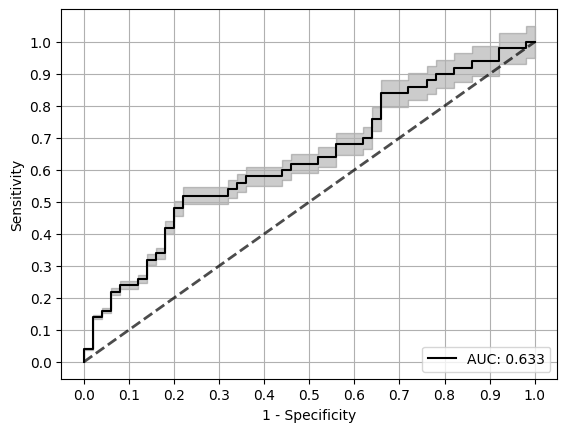}
\label{fig1}
\end{figure}

\begin{figure}[!ht]
\caption{{\bf ROC Curve for the baseline (3 Biomarkers).}
}
\includegraphics[scale=0.6]{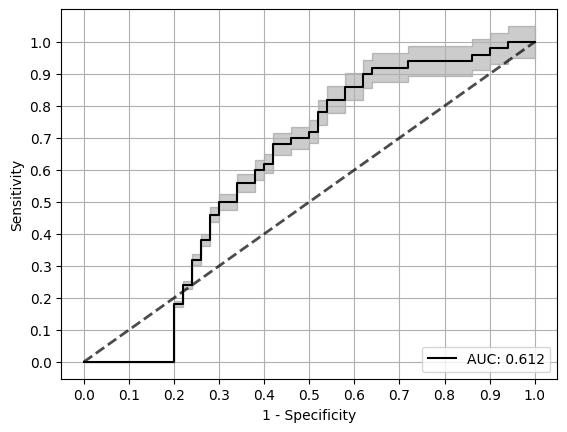}
\label{fig2}
\end{figure}

Overall, the most performance in terms of test AUC was observed for the deep neural multilayer perceptron (MLP) classifier with causal metric for biomarker selection, which outperformed the baseline by $0.114$, shown in Table~\ref{table2}. For Sensitivity at specificity of $0.9$, XGB with causal metric (as seen in figure~\ref{fig1}) outperformed the baseline (as seen in figure~\ref{fig2}) by $0.240$. Notably, the use of causality feature selection improved performance irrespective of classifier, providing a minimum improvement of $0.120$ (binarized) over univariate feature selection for each classifier for $Sen@90$ (Table~\ref{table2}). Comparable results were noted for Sensitivity when Specificity was set to $0.8$ along with test AUC. 

We note that training AUC was strongest for random forest with univariate selection with a value of $0.997$ – however, this drops to $0.558$ for testing.  This is surprising, as random forest generally does not overfit (Breiman~\cite{bib27}) however it may indicate that univariate feature selection may cause overfitting when used in more complex models – as we observed the large discrepancies between training and testing AUCs when univariate feature selection was used in all cases except logistic regression. On the other hand, the average drop for the causality measure is $0.118$ and a maximum of $0.186$ while there is an average drop of $0.260$ and a maximum of $0.439$ for univariate feature selection which indicates a possibility of overfitting caused when causality is ablated.

\begin{table}[!ht]
\centering
\caption{
{\bf Results for 3 biomarkers using 5 models with causal-based and univariate feature selection }}
\begin{tabular}{|l+l|l|l|l|l|l|} 
    \hline
    \textbf{Model} & Method & Train AUC & Test AUC & $Sen@90$ & $Sen@80$ \\ 
    \thickhline 
    \textbf{MLP}& Univariate & 0.937 & 0.581 & 0.080 & 0.140\\
    \hline & Univariate(B)& 0.738& 0.527& 0.000& 0.000 \\
    \hline & Causal& 0.720& {\ul \textbf{0.695}}& 0.220& 0.420\\
    \hline & Causal(B)& 0.774& 0.588& 0.200& 0.300\\ 
    \hline\textbf{XGB}&Univariate&0.969&0.613&0.200&0.260\\
    \hline&Univariate(B)&0.754&0.538&0.000&0.000\\
    \hline&{\ul\textbf{Causal}}&{\ul\textbf{0.719}}&{\ul\textbf{0.633}}&{\ul\textbf{0.240}}&{\ul\textbf{0.480}}\\
    \hline&Causal(B)&0.611&0.463&0.200&0.340\\
    \hline\textbf{LR}&Univariate&0.699&0.612&0.000&0.180\\
    \hline&Univariate(B)&0.756&0.560&0.000&0.000\\
    \hline\textbf{}&Causal&0.678&0.510&0.180&0.280\\
    \hline&Causal(B)&0.771&0.594&0.200&0.200\\
    \hline\textbf{GBT}&Univariate&0.984&0.571&0.120&0.280\\
    \hline&Univariate(B)&0.738&0.527&0.000&0.000\\
    \hline\textbf{}&Causal&0.722&0.659&0.140&0.420\\
    \hline&Causal(B)&0.613&0.496&0.220&0.360\\
    \hline\textbf{RF}&Univariate&{\ul\textbf{0.997}}&0.558&0.120&0.200\\
    \hline&Univariate(B)&0.736&0.620&0.060&0.080\\
    \hline&Causal&0.719&0.593&0.120&0.120\\
    \hline\textbf{}&Causal(B)&0.662&0.583&0.180&{\ul\textbf{0.540}}\\
 \hline
\end{tabular}
\label{table2}
\caption*{(B) dictates using binarized data; \textbf{Bolded} values dictate better\\
performance; \underline{\textbf{Underlined}} values dictate best performance}
\end{table}



\subsection*{Selection of 10 Biomarkers}

\begin{figure}[!ht]
\caption{{\bf ROC Curve for GBT model with univariate feature selection (10 Biomarkers).}
}
\includegraphics[scale=0.6]{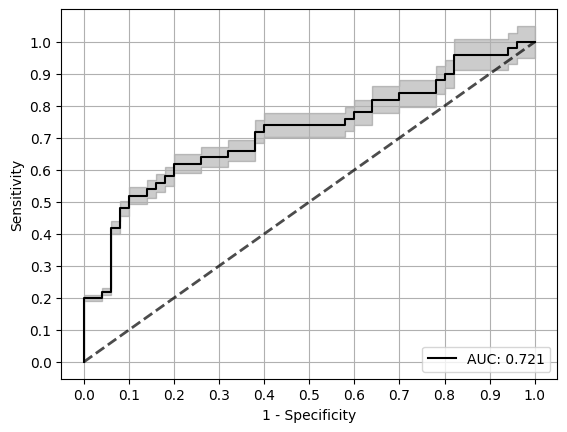}
\label{fig3}
\end{figure}

\begin{figure}[!ht]
\caption{{\bf ROC Curve for the Baseline (10 Biomarkers).}
}
\includegraphics[scale=0.6]{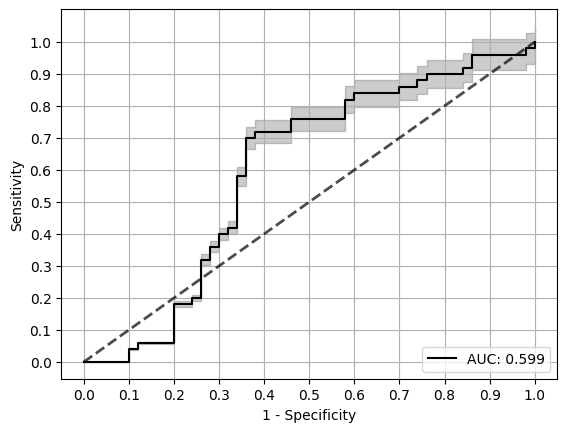}
\label{fig4}
\end{figure}

On the other hand, the best-performing model, with respect to test AUC, was MLP with univariate feature selection, which outperformed MLP with causality measure by $0.286$, shown in Table~\ref{table3}. Furthermore, GBT with univariate feature selection (as seen in figure~\ref{fig3}) reported the highest sensitivity at a specificity of $0.9$, that is $0.520$ while GBT with causality measure reported sensitivity at a specificity of $0.9$ as $0.22$. Also, the baseline (as seen in figure~\ref{fig4}) gave a moderate test AUC of $0.599$ but a low $Sen@90$ value.
We found that, with a high number of biomarkers, univariate feature selection seems to be performing well with respect to test AUC compared to the causality measure for all methods by a minimum of $0.025$ (binarized) and $0.029$ (non-binarized). 

For a higher number of biomarkers, a more generic method like univariate seems to suffice. While increasing the historical data might help improve the performance of other approaches, the less data-hungry causal approach already performs well without inconsistent sensitivity at a specificity of $0.9, 0.8$.

\begin{table}[!ht]
\centering
\caption{
{\bf Results for 10 biomarkers using 5 models with causal-based and univariate feature selection
 }}
\begin{tabular}{|l+l|l|l|l|l|l|} 
    \hline
    \textbf{Model} & Method & Train AUC & Test AUC & $Sen@90$ & $Sen@80$ \\ 
    \thickhline\textbf{MLP}&Univariate&1.000&0.669&0.140&0.340\\
    \hline&Univariate(B)&\textbf{0.926}&\textbf{0.764}&\textbf{0.480}&\textbf{0.480}\\
    \hline\textbf{}&Causal&0.909&0.551&0.200&0.260\\
    \hline&Causal(B)&0.918&0.478&0.120&0.200\\
    \hline\textbf{XGB}&Univariate&0.998&0.701&0.300&0.420\\
    \hline&Univariate(B)&0.890&0.684&0.460&0.460\\
    \hline\textbf{}&Causal&0.816&0.575&0.200&0.340\\
    \hline&Causal(B)&0.879&0.659&0.220&0.360\\
    \hline\textbf{LR}&Univariate&0.811&0.599&0.040&0.180\\
    \hline&Univariate(B)&0.878&0.746&0.460&0.480\\
    \hline\textbf{}&Causal&0.734&0.569&0.080&0.220\\
    \hline&Causal(B)&0.830&0.681&0.320&0.360\\
    \hline\textbf{GBT}&{\ul\textbf{Univariate}}&{\ul\textbf{1.000}}&{\ul\textbf{0.721}}&{\ul\textbf{0.520}}&{\ul\textbf{0.620}}\\
    \hline&Univariate(B)&0.919&0.746&0.480&0.500\\
    \hline\textbf{}&Causal&0.852&0.588&0.220&0.260\\
    \hline&Causal(B)&0.875&0.540&0.180&0.220\\
    \hline\textbf{RF}&Univariate&{{\ul\textbf{0.999}}}&0.649&0.140&0.380\\
    \hline&Univariate(B)&0.926&0.708&0.420&0.440\\
    \hline&Causal&0.894&0.594&0.140&0.260\\
    \hline\textbf{}&Causal(B)&0.904&0.538&0.120&0.320\\
 \hline
\end{tabular}
\label{table3}
\caption*{(B) dictates using binarized data; \textbf{Bolded} values dictate better\\
performance; \underline{\textbf{Underlined}} values dictate best performance}
\end{table}

\subsection*{Hyperparameter Study}
\label{sec: hyper}
\begin{figure}[!h]
    \centering
    \begin{subfigure}{0.49\textwidth}
        \caption{ }
        \includegraphics[width=0.9\linewidth]
        {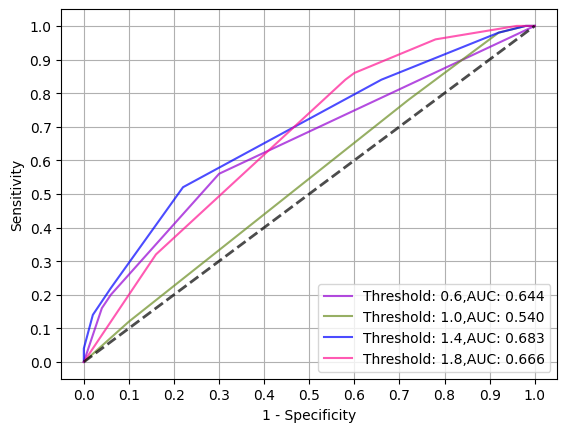}
        \label{fig5:a}
    \end{subfigure}
    \hfill
    \begin{subfigure}{0.49\textwidth}
        \caption{}
        \includegraphics[width=0.9\linewidth]{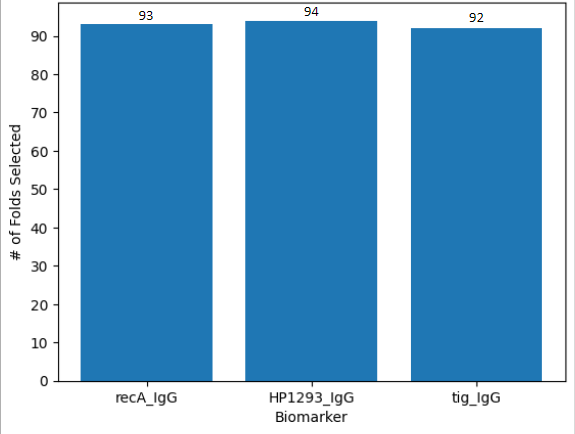}
        \label{fig5:b}
    \end{subfigure}
    \hfill
        \begin{subfigure}{0.49\textwidth}
        \caption{}
        \includegraphics[width=0.9\linewidth]{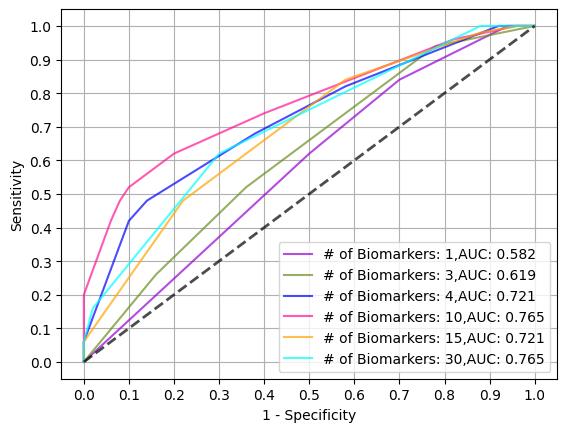}
        \label{fig5:c}
    \end{subfigure}
\caption{{\bf Hyperparameter Sensitivity} a: ROC Curve with multiple thresholds($\gamma$) for XGB  model with causal-based biomarker selection. b: Frequency of Selected Biomarkers, where $K=3$. c: Effect of $K$ for threshold $1.4$ for GBT model with univariate selection.
}
\label{fig5}
\end{figure}

As shown in Table~\ref{table2} and Table~\ref{table3}, a few methods were classified based on the binarization of biomarker values before model training indicated by B; for example, Causal(B) means causality method with binarized inputs. We discretize all input measurements for a given sample based on a threshold $\gamma \in \{0.6, 1.0, 1.4, 1.8\}$. The tables ~\ref{table2} and ~\ref{table3} are for threshold value $\gamma = 1.4$.  However, it is important to note that there is little variance in AUCs for most thresholds, showing the stability of the selected biomarkers as seen in Fig~\ref{fig5:a}. Also, consistency is observed in the frequency of biomarker selection. Furthermore, by raising the value of $K$ significantly, we get diminishing returns, suggesting a saturation point to pick the number of biomarkers, $K$.\\

We found the biomarkers: DNA-directed RNA polymerase subunit alpha $HP1293$, recombinase RecA $recA$, and trigger factor $tig$ IgG antibodies, to be the most frequently selected biomarkers related to gastric cancer. Fig~\ref{fig5:b} shows the high frequency of biomarkers $recA IgG, HP1293 IgG,$ and $tig IgG$, appearing in above $90\%$ of all folds when evaluating with LOOCV, therefore supporting the stability of the model. These are the biomarkers that were consistently picked by the causality measure.

Notably, the test AUC increases with $K$ and saturates after $K=10$ as seen in Fig~\ref{fig5:c}. However, $K$ had a limited impact on the biologically relevant $Sen@90$ measure.  Initially, increasing the value of $K$ increased the test AUC by the magnitude of $0.2$. As we gradually increased $K$, the test AUC levels out to a certain range, around $0.7$ but the $Sen@90$ measure tends to get more sparse. We see diminishing returns by adding any more number of biomarkers. This relation has relevance based on the target application desired to make inexpensive diagnostic kits. 

\section*{Discussion}
 We see the effects of ablating away causality measure with univariate feature selection in Table \ref{table2}. We observe a higher AUC and consistent sensitivity values for the causal method as we decrease the number of biomarkers, and these benefits go away otherwise. This will be beneficial when applying this method to the industry considering the computational power and being less expensive as the method performs better with less number of biomarkers. This approach can also be applicable to similar domains for other disease prediction. Additionally, the experiments with the causal metric can be extended by adding a combinatorial way of picking the ranked causal biomarkers.

\section*{Conclusion}
In this paper, we use a causality measure to select biomarkers paired with ML-based classifiers on a gastric cancer dataset for disease detection purposes. We pre-select biomarkers to reduce the number of biomarkers considered to be more practical, reduce overfitting, and to understand the causal effect of the set of biomarkers.
With respect to $Sen@90$, and $Sen@80$, the XGB model with causality measure performed better when compared to the baseline for $3$ biomarkers and has a hike of $0.114$ on AUC. We found that approaches with the causal metric performed better when handling a smaller number of biomarkers, while conventional techniques like univariate feature selection performed better with a larger number of biomarkers. 
The causality measure compares co-occurring biomarkers, they could provide biological intuition enabling further empirical studies. We see evidence that this approach likely generalizes for the prediction of other diseases based on biomarkers, as our machine learning methods perform well across a variety of diseases.

\nolinenumbers

\bibliographystyle{vancouver}
\appendix

\section*{Supporting information}
 \subsection*{S1 Appendix}{\bf Learning models}
 \label{S1_Appendix}
    \paragraph*{Multi layer perceptron} 
    A multi-layer perceptron (MLP) is a type of artificial neural network (Gardner et al.~\cite{bib22}). MLPs are composed of multiple layers of interconnected nodes or neurons, each of which performs a non-linear operation on the input data. The input layer of an MLP receives the input data, and the output layer produces the predicted output. In between the input and output layers, there can be one or more hidden layers, each of which contains multiple neurons. The neurons in each layer are connected to the neurons in the next layer, forming a dense, fully connected network. During training, an MLP adjusts the weights of the connections between neurons to minimize the difference between the predicted output and the actual output. This process is performed using a process called backpropagation, which involves propagating the error through the network and updating the weights using gradient descent.
    \paragraph*{Extreme Gradient Boosting/XGBoost} 
    XGBoost (Chen et al.~\cite{bib28}) is designed to improve the performance of gradient-boosted trees, especially in terms of speed and model accuracy. Like other gradient-boosting algorithms, XGBoost builds an ensemble of decision trees to make predictions. However, XGBoost introduces several improvements to the gradient boosting algorithm, such as a novel tree construction algorithm, parallel processing, and a regularized learning objective.
    \paragraph*{Logistic regression} 
    It is a type of regression analysis that predicts the probability of an outcome based on one or more predictor variables (Kleinbaum et al.~\cite{bib20}). In logistic regression, the dependent variable is binary, meaning it can take only two possible values. The algorithm models the relationship between the independent variables and the binary outcome using the logistic function, also known as the sigmoid function. The output of the sigmoid function is a value between 0 and 1, which represents the predicted probability of the positive class.
    \paragraph*{Gradient-boosted trees}
    Gradient-boosted trees combine the predictions of multiple decision trees to improve the accuracy of predictions. In gradient-boosted trees, decision trees are created in a sequence, and each new tree is built to correct the errors of the previous tree. The algorithm assigns more weight to the data points that were incorrectly predicted in the previous iteration, and less weight to the correctly predicted points. This process continues until the algorithm reaches a predefined stopping point, such as a maximum number of iterations, or when the accuracy of the model stops improving.
    \paragraph*{Random forest} 
    It is an ensemble learning method that combines multiple decision trees to make predictions (Breiman et al.~\cite{bib27}). Unlike a single decision tree, Random Forest creates multiple decision trees on randomly selected subsets of the training data and randomly selected subsets of the features. Each decision tree in the forest is constructed independently, and the final prediction is made by combining the predictions made by all the trees in the forest, typically by taking the average or the majority vote of the predictions.
    \paragraph*{Univariate Selection (UV)} 
    A standard Logistic Regression model with Univariate Feature Selection was used as the baseline. This baseline model enabled us to identify the most relevant features by selecting $K$ features, where $K$ was set to $3$ and $10$. This approach provided us with a reference point to compare the performance of our model.
    The hyperparameter values used for our experiments for each model is mentioned in table~\ref{s1table}
    \renewcommand{\thetable}{S1}
    \begin{raggedright}
    \begin{table}[!ht]
    \centering
    \caption{\textbf{Hyperparameters used for each model}}
    \begin{tabular}{|l l+l l|l|}
     \hline
     \textbf{MLP}& & \textbf{XGB, GBT} &  \\ 
     \hline
     hidden\_layer\_sizes:& $256, 128, 64, 32$ & max\_depth: & 2.0 \\
     activation:&relu&learning\_rate: & 1.0 \\
     
     random\_state:&1.0&n\_estimators: & 10.0 \\
      & &random\_state: & 0.0 \\ 
       \thickhline
     \textbf{LR}& & \textbf{RF} &\\ 
        \hline
        solver: &lbfgs& n\_estimators:& 10.0 \\
    
     \hline
    \end{tabular}
    \label{s1table}
    \end{table}
    \end{raggedright}

\subsection*{S2 Appendix}{\bf Derivation of Causal Metric}
\label{S2_Appendix}

 Our method, the process of selecting the top $K$ biomarkers, uses only the training data. Here is an overview:
 
\begin{enumerate}
\item For each individual biomarker, we compute the sensitivity and specificity with respect to the training data by classifying samples solely on if the associated biomarker reading for that single biomarker exceeds threshold $\gamma$.
\item For each biomarker $i$, we compute $s2$ metric value which is simply the specificity multiplied by the sensitivity calculated based on a threshold $\gamma$ in the above step, giving us the $s2$ metric value for a biomarker $i$.
\item Using the causal computation (Equation~\ref{eqn:cmetric}) we compute $causal_\gamma(i)$ for every biomarker.  We then rank all biomarkers by this metric and select the top $K$.
\end{enumerate}

Here are further technical details on the derivation of causal metric to rank biomarkers:

 For hyperparameter $\gamma$ we will use the notation $X^{(\gamma)}$ and $x_i^{(\gamma)}$ to be a matrix (or vector respectively) consisting of zeros and ones based on the real-valued threshold $\gamma$ (values are set to $1$ if the feature value is greater than or equal to $\gamma$). For a given feature vector $x_i^{(\gamma)}$, we will use the notation $Spec(x_i^{(\gamma)})$ and $Sens(x_i^{(\gamma)})$ to be the specificity and sensitivity if only that binarized feature vector is used to make a prediction (in other words, we predict each sample of $y$ if the value for feature $i$ exceeds $\gamma$).  We will use the notation:

\begin{eqnarray}
\label{eqn:s2}
s2(x_i^{(\gamma)})=Spec(x_i^{(\gamma)}) \times Sens(x_i^{(\gamma)})
\end{eqnarray}

Sensitivity and specificity measures are standard measures used in bio-medicine. We use $s2$ measure to indicate the average influence of the presence or lack of a biomarker for samples with the disease, and the influence of the presence of a biomarker on samples without the disease. We examine the average increase of this causal effect of the biomarkers on the disease based on all possible co-occurring biomarkers.

We will consider the samples whose $s2$ metric value is greater than the average $s2$ metric value. Consider all biomarkers $R_i$ to be related biomarkers for $i$ when there is an overlap of at least one sample between the subset of case samples where the biomarker value exceeds the threshold. In other words, for a given biomarker $i$, we say has a value of 1. $R_i$ is all of the other biomarkers that have a value of 1 when the sample is of a case. 

We will use the logical operations of negation, and disjunction to take binarized vectors and form new ones, and this follows the normal intuition.  Technical definitions are defined below:
\begin{itemize}
\item \textit{Negation.} For a given binarized vector $x_i^{(\gamma)}$, we define $not_\gamma(i)$ where each component equals one minus the corresponding component of $x_i^{(\gamma)}$ (i.e., the zeros and ones are switched).  When refering to feautres by their index (i.e., $j$) we will use the notation $\neg j$ to refer to the ``index'' pointing to vector $not_\gamma(j)$.
\item \textit{Disjunction.}  For two binarized vectors $x_i^{(\gamma)}, x_j^{(\gamma)}$ we define $disj_\gamma(i,j)$ where the $k$th position is equal to $\min(1,x_i^{(\gamma)}[k]+x_j^{(\gamma)}[k])$.  In other words, it  is a vector (of the same size of both inputs) where each position is the sum of the pairwise components in each vector clipped to $1$ (i.e., each component is $1$ if either the corresponding component in $x_i^{(\gamma)}$ or $ x_j^{(\gamma)}$ is $1$ and zero otherwise).

\end{itemize}
Each hyperparameter $\gamma$ will have different related biomarker sets. For a given vector $x_i^{(\gamma)}$, the set of related biomarkers for the threshold $\gamma$, $R_{\gamma}(i)$ is the set of all other vector indices $j$ such that $\sum disj_\gamma(i,j)>0$.

We now introduce a \textit{disjunctive causal ranking metric} for features.  It is defined as follows:

\begin{eqnarray}
\label{eqn:genmetric}
causal(i)=\frac{\sum_{j \in R_i}f(i,j)-f(\neg i, j)}{size(R_i)}
\end{eqnarray}

Intuitively, $causal_\gamma(i)$ tells us the average increase in the $s2$ metric obtained when feature $i$ is used vs. when it is not.
We compute this metric in training for all features and then, based on hyperparameter $k$, we select the $k$ features with the greatest value for $causal(i)$ and use those to train the model.  Note that the samples used for computing each  $causal(i)$ and training the model are the same.
This is used to identify the causal factors that have the most effect on the model to make a decision as well as not consider the factors that make a small difference.

\begin{eqnarray}
\label{eqn:cmetric}
causal_\gamma(i)=\frac{\sum_{j \in R_\gamma(i)}s2(disj_\gamma(i,j))-s2(disj_\gamma(\neg i, j))}{size(R_\gamma(i))}
\end{eqnarray}

 We rank the biomarkers based on the causal measure. Assuming there is a certain combination of biomarkers $B_1, B_2, B_3$ consistently present for the class of cancer, they have approximately the same values, outcome implies they are associated with the same causal measure. When we pick biomarkers with respect to the order of causality, we would pick all those three biomarkers. $B_1, B_2, B_3$ will have the same causal measure and the probability of a sample having cancer which showed $B_1$ will be the same as $ B_2, B_3$ and lack of presence of any of these biomarkers will not contribute to the prediction of cancer or not. 

\end{document}